\documentclass[journal]{IEEEtran}
\usepackage{amsmath}

\usepackage{cite}

 \usepackage{graphicx}
\usepackage{cite}
\usepackage{xcolor}

\ifCLASSINFOpdf
\else
\fi
\begin{document}
%
\title{Variability-aware Memristive Crossbars - A Tutorial}
%
%
%

\author{A.P. James,~\IEEEmembership{Fellow,~IET,}
                and~L.O. Chua,~\IEEEmembership{Fellow,~IEEE}
}

%
%

\markboth{}%
{}
%



\maketitle

\begin{abstract}
Memristor crossbar architecture is one of the most popular circuit configurations due to its wide range of practical applications.  The crossbar architecture can emulate the weighted summation operation, called multiply and accumulate operation (MAC). The errors to MAC computing get introduced due to a range of crossbar variability. We broadly group the variability in three categories: (1) device-to-device variations, (2) programming nonlinearity, and (3) those from peripheral  circuits. This tutorial provides insights into the variability and compensation approaches that can be adopted to reduce its impact when designing for practical applications with crossbars. .
\end{abstract}


%
\IEEEpeerreviewmaketitle

\section{Introduction}
%
%
%
%


\IEEEPARstart{M}{emristive} systems are abundant in nature, with many biological and natural systems exhibiting memristive properties. The idealistic models in Fig. 1, can be used for understanding and laying the foundation dynamic behavior of memristor using dynamic route maps (DRM) and $i-v$ plots \cite{leon2015everything,chua2018five,ascoli2016history}. The DRM in the phase plane can determine the optimal programming pulse height and width along the identified equilibrium points.  While these equations can describe the devices in idealistic situations, experimentally, it is harder to validate under noisy conditions. In contrast,  \textit{Coincident Zero-Crossing Signatures} can be experimentally observed in many devices. {The ideal models do not consider the impact of noises or the variability that practical devices show, and the development of generic  \textit{imperfect memristor} models \cite{leon2015everything, linn2014applicability} remains an open problem.}

\begin{figure}[!ht]
    \centering
    \includegraphics[width=80mm]{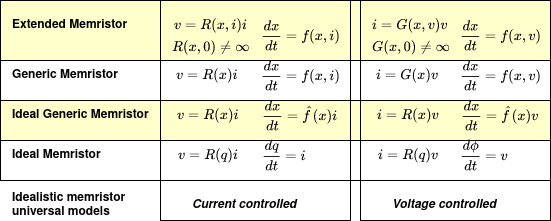}
    \caption{Universe of memristor models, excluding the parasitic and other device variability. }
    \label{fig:my_label}
\end{figure}

Several devices show the resistive switching behavior having nonvolatile properties, grouped under the board class of memristive devices. This classification assumes that not all of these devices show idealistic properties of the  memristor but have many properties of ideal memristor that can be used for building practical memristive systems \cite{zidan2018future, jimenez2021compact}. For example, nonvolatility, switching between conductance values, small form factors, integration with CMOS, and synaptic behavior, are all attractive properties useful for building emerging on-chip applications. Emulating neural networks, analog computing, and in-memory computing are growing applications of memristive systems. 

Crossbar arrangement of memristor \cite{vourkas2012novel, xia2019memristive,li2021memristive} is a popular approach to build memristive computing applications. Any errors in the crossbar will impact the overall accuracy of computations\cite{chen2021multiply}. Such errors develop due to the variability of devices, circuit parasitic, and device aging\cite{gao2021memristor}. For ensuring reliable use of memristive crossbars, it is important to design the overall system so that variability is compensated or reduced to practical limits allowable for a given application. The attempts to include variability in simulations \cite{reuben2020incorporating, ostrovskii2022structural, adam2021generalised} are important for accurate analysis of circuit performance.

{This tutorial draws motivation from the above  mentioned emerging design and performance challenges of building memristor crossbar based systems. So far, the previous works \cite{zidan2018future,li2018review, xia2019memristive,li2021memristive} in the domain largely focuses on the applications and less on the issues of variability.  This tutorial contributes to give a concise account of the types of variability, its implications, and possible directions to create a variability-aware crossbar in the practical design of memristive applications.  } Section II gives an overview of the crossbar, while Section III provides insights on variability and compensation approaches, and Section IV provides the summary.

\section{Crossbar overview}

The crossbar consists of memristors connected in a matrix arrangement, as shown in Fig. 2, with multiple inputs and outputs. The inputs are applied along the rows, and outputs are read along the columns.

\begin{figure}
    \centering
    \includegraphics[width=70mm]{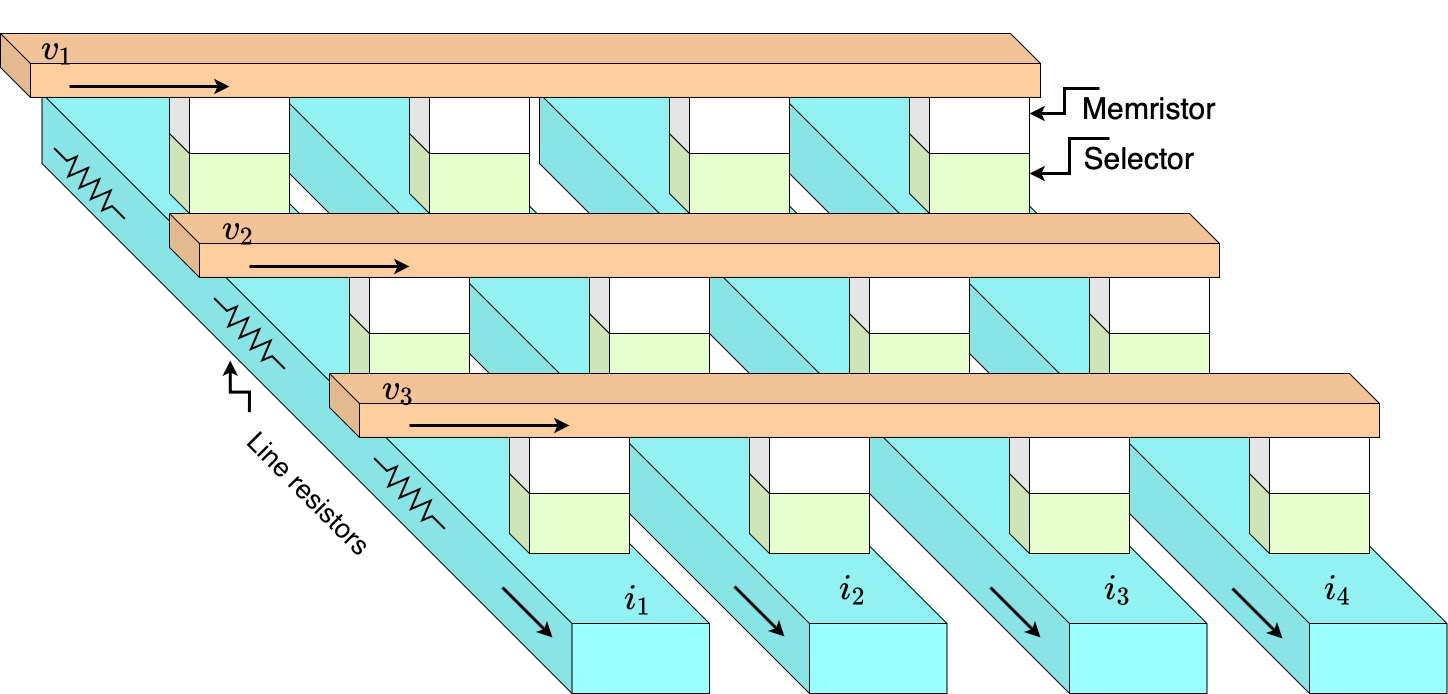}
    \caption{Example of memristor crossbar of array size 3 $\times$ 4. Each crossbar node has a memristor and selector device.}
    \label{fig:my_label}
\end{figure}

\subsection{Crossbar structure}

The most popular memristor structure is the crossbar with inputs as voltages $v_m$ and outputs as currents $i_n$. The nodes of the crossbar consist of a memristor and selector device (e.g., transistor or diode), having a conductance of $G_{mn}$. The selector devices are required to avoid the sneak path currents from the neighborhood columns.
The overall column current $i_n= \sum_{m=1}^M G_{mn}v_m$, is equivalent to the weighted summation operation, called multiply and accumulate (MAC) operation in analog computations. 

\begin{figure}[!ht]
    \centering
    \includegraphics[width=60mm]{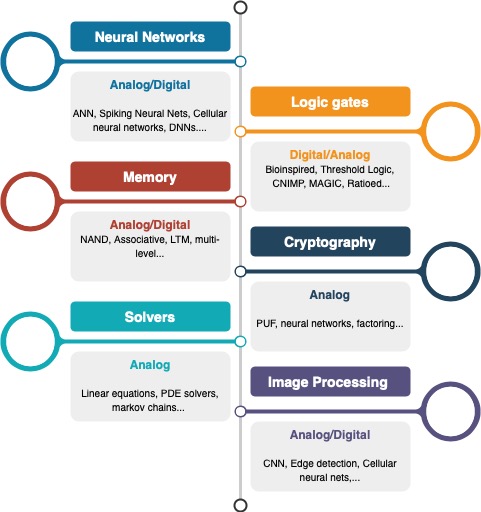}
    \caption{Examples of major applications using crossbar arrays.}
    \label{fig:my_label}
\end{figure}

\subsection{Crossbar applications}

The analog MAC operation is the hallmark of the crossbar, useful for a range of applications.  {For example, in neural computations, for $M$ inputs $y_m$ with weights $w_m$, the output of neuron is expressed as $\sum_{m=1}^M w_{m}y_m$, which maps with each column current outputs of the crossbar. Thus, each crossbar column can be considered as partial neuron computations, and the crossbar as a single neuron layer. } The variability in the crossbar can be problematic for some applications while useful for others. 
Figure 3 shows the broad classification of major crossbar applications. Neural networks \cite{li2018review} such as Spiking neural networks \cite{mehonic2020memristors}, Hierarchical Memory Networks \cite{krestinskaya2018hierarchical}, or Deep Learning Networks \cite{mehonic2020memristors} can be implemented in both analog/digital crossbar circuits, with a range of variability using the neural training schemes. Application of logic computing is also robust to variability due to the binary logic, while bit errors are possible when there are thresholds to compare or are stuck at fault errors. The use of crossbar for binary or analog/discrete memory with high density 2D and 3D crossbar arrays could be limited in readout speeds with increased variability. The crossbar variability can be used to build unique physically unclonable functions (PUFs) useful for a variety of cryptographic algorithms \cite{james2019overview}. The crossbar analog MAC computations can be used for building solvers such as for PDE\cite{zidan2018general}, discrete Markov chains \cite{zoppo2021analog}, or linear program solvers\cite{cai2016low}, where the accuracy of analog MAC becomes crucial in controlling with increasing variability. Image processing\cite{li2018analogue} is another popular application, with crossbar useful for edge detection\cite{pannu2020design}, face detection\cite{james2017htm}, and object detection, where the accuracy and speed of detection is sensitive to aging and variability.

\begin{figure}[!ht]
    \centering
    \includegraphics[width=70mm]{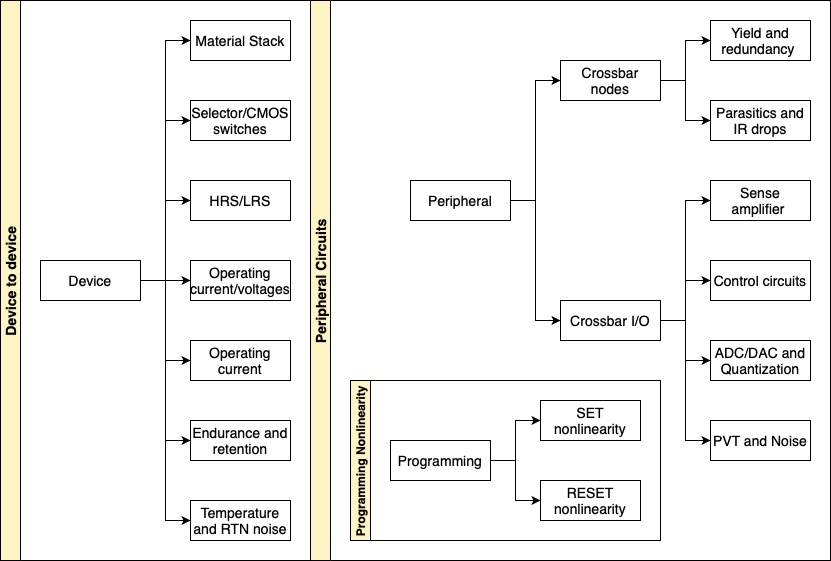}
    \caption{The broad classification of the types of the variability faced in memristor crossbars.}
    \label{fig:my_label}
\end{figure}

\section{Crossbar Variability and Compensation}

The definition of the variability in memristive crossbar systems varies from the device level to the hardware system level. The inaccuracy of the MAC computation can significantly impact the reliability of using crossbar in various applications. Figure 4 shows a broad classification of variability in a memristive crossbar.

\subsection{Device-to-device variability}

Most of the memristive devices used for building crossbars are in their experimental stage. For commercial use, a new device is expected to have a low  device-to-device variability \cite{li2018review,nickel2013memristor, laube2021device, kumar2021uncertainty} and less unpredictable output current errors. The main design considerations include the following:
\begin{enumerate}
    \item {\textit{Material stack and
switching physics.}} The material stack has a direct role in the thermal effects, chemical reactions, ionic transfers, spin polarization, and phase transition of resistive switching. The variations in material stacks due to defects and unstable manufacturing process leads to non-uniform resistive switching between the devices.
\item  {\textit{CMOS technology node.}} CMOS transistors used in crossbar nodes and at the inputs or outputs of crossbar usually have varying levels of technology-specific leakage currents and parasitic effects that impact the dynamic response times of the crossbar column computations. 

\item  {$R_{on}/R_{off}$ \textit{ratio}.} Changes in ratio (low-resistance state (LRS) $R_{on}$ and high-resistance state (HRS) $R_{off}$) can complicate the programing of memristors and increase the relative current errors. This can   introduce changes to dynamic switching behavior expected for a certain frequency of programming pulses, which require readjusting the pulse frequency and amplitude from one device to another to achieve the same state outcomes. 

\item {\textit{Range of operating currents from devices}.} The write, read and erase currents in a memristor need to be much higher than sneak path currents, while a larger current would increase the power dissipation in the crossbar. Optimizing the operating range of current memristors for power, density, and accuracy is essential to ensure the practical usefulness of the crossbar.

\item {\textit{Read and write voltage ranges.}} Lower voltages would imply the need for longer pulses to program the memristors, while higher voltages would increase the power dissipation in the crossbar. The variation in voltages that the memristor supports, if varied between the devices, introduces uneven distribution of power within the crossbar leading to different stress levels on the devices.

\item {\textit{Endurance and retention.}} The four major memristor mechanisms are based on a redox reaction, phase change, magnetic polarization, and ferroelectric polarization. In each of these device types, endurance varies from 10$^{9}$ to 10$^{14}$, and retention is more than ten years. Any reduction in endurance implies a reduced lifespan in applications involving frequent updates of conductance. Reduction in retention times makes the crossbar unreliable as a memory or for in-memory computing applications, such as neural networks.

\item {\textit{Temperature dependence.}} The temperature can influence the flow of current in the semiconductor material. The memristor current increases with an increase in temperature while switching delays decrease. The temperature change from 200k to 400k increases the currents by an order of two, implying higher power dissipation during crossbar computations. The changes in switching delays with varying temperatures can make memristor programming unreliable.

\item {\textit{Random telegraph noise (RTN).}} RTN with current signals is observed in memristors made of metal-insulator-metal. The nature of the RTN signals is influenced by materials used for electrodes and insulators, type of deposition method, and insulator thickness. 

\end{enumerate}

The variations in these parameters result in MAC computation errors. It has been shown that device-to-device variation in phase-change random access memory arrays is more difficult to train than devices exhibiting consistent cycle-to-cycle variation.  This indicates that any reduction in the device-to-device variability in practical applications can improve the overall application reliability.

\subsection{Programming nonlinearity}

In many memristors such as HfO$_2$ or PCM, the memristance can be represented as a nonlinear function of time \cite{ambrogio2018equivalent,wu2018methodology}.
Both linear and nonlinear drift models \cite{mcdonald2010analysis} can be used for modeling the memristance.  In contrast, if memristance variation is linear in time, their programming is easier as memristance is proportional to the width of a programming voltage pulse. 

In practical realization of memristive devices, such as analog RRAMs, linear adjustment (tuning) of conductance is not possible with a sequence of identical pulses\cite{wu2018methodology}. Both the SET and {RESET} cycles observe nonlinearity and require device level modifications or circuit level node design to ensure linearity.

A device level solution has been demonstrated in the past by linearly setting SET and {RESET}  using  HfOx type memristors by adding electro-thermal
modulation layer to the switching layer to control the dynamics of the conductance changes. The circuit level solution includes an anti-serial architecture by connecting two memristors of opposite polarities. The complementary action of the serially connected memristors in such configuration shows linear behavior in time, useful for linearly programming the crossbar node.

\subsection{Peripheral circuits}

\subsubsection{Crossbar circuits}
Several interface circuits are required to make crossbars useful for practical applications.
The crossbar array is directly influenced by variations in yield, redundancy, circuit parasitics, current-resistance (IR) drop, and
array size.

\paragraph{Yield and redundancy} High yield and redundancy are required for the use of crossbars as memories. Redundancy is essential in applications where data recovery becomes important \cite{kannan2014detection, shen2021variability}, and in-memory computing should be highly accurate.

\paragraph{Circuit parasitics} 
The circuit parasitics\cite{itoh2016parasitic,jeong2017parasitic} include the effects of wire resistance\cite{zhang2021impact} from metal lines connecting the memristors, stray capacitors, and resistors in the CMOS switches, creating RC delays along the crossbar lines. The impact is more pronounced as the size of the crossbar array increases, reducing the read speeds of the system. The parasitic coupled with large device-to-device variations in the crossbar can eventually lead to high relative current errors making real-time applications unreliable.    

\paragraph{IR drops} Both static and dynamic IR drops happen in a memristor crossbar. The voltage IR-drop \cite{liu2014reduction} results from the resistance of the crossbar arrays and metal wires. The node resistance $G$ becomes a function of parasitic resistance of switches, memristance, and wire resistance. The variability in memristor devices, wires, or switches will impact the 1T1M crossbar read and write accuracy. The IR drops, if not compensated in write stages by readjusting the conductance values, will also impact the readout values.  

Compensation by readjusting the programming pulse amplitude and width can effectively reduce the IR drop impact. Suppose $T$ is the programming pulse widths, and $V$ is the pulse amplitude for programming the crossbar without considering IR drops. When IR drops are considered, the voltage distribution $V$ gets distorted to $V'$. For compensating for the drop in amplitude, the pulse width for $V'$ is reworked to get a new time period $T'$ that gives the desired state under IR drops.

\subsubsection{Crossbar I/O}

The I/O blocks to the crossbar often involve a variety of analog or digital circuits. The most common circuits are OpAmps and sense amplifiers, line selector switches, programming circuits, data converters (ADCs and DACs), buffers, and multiplexing circuits. The variability and reliability of these circuit elements determine the practical feasibility of using crossbars in a range of applications.

\paragraph{Sense amplifiers}
Sense amplifiers are required to readout the column currents. In analog neural networks designed with crossbars, the difference between two column currents is used to account for the  negative weights of the networks. In most other applications, one column current output is readout using an opamp or set of amplifiers. The opamps are expected to have low input offsets (current and voltage), low delays, and work in high frequencies. The parasitic capacitors and resistances from the sense amplifiers can uneven delays in parallel column reads of the crossbar, requiring timing and memory controls before passing the signals to the subsequent crossbars or circuits. 

\paragraph{Control circuits}

Different sets of control circuits are required to control various switches within and on the peripherals of the crossbar. For example, the most commonly used 1T1M crossbar configuration will require control signals applied to the gate of the transistor and input/output lines for the read, write, and erase operations. In noise-free control signals, accurate timing is essential for programing the memristors in the crossbar. Further, in neural networks with multiple crossbars, control circuits are required to access the intermediate storage and multiplex the signals across multiple crossbars, such as in tiled crossbar configuration.

\paragraph{ADC/DAC} ADC is required to convert the analog column outputs from the crossbar to digital for further processing and storage. DAC is required for converting the digital input signal to analog for use in binary crossbar arrays. The precision of ADC/DAC and the ADC multiplexing would form the primary design considerations. The common errors in ADC include quantization error, offset error, gain error, and nonlinearity, all of which impact the overall use of crossbar for applications. The absolute errors from the ADC can lead to increased output current errors, which, when implemented in applications such as neural networks, leads to reduced inference accuracy.     

\paragraph{Binary crossbar and quantization} If the crossbar is used for storing binary weights as in binary neural networks, errors from partial sum quantization are also possible. Any inaccuracy in DAC and ADC further degrades the errors due to partial sum quantization.   

\paragraph{Process, voltage and temperature (PVT), and noise} The peripheral circuits can be PVT sensitive and induce noise in the crossbar circuits. The PVT sensitivity of the peripheral circuits can limit the practical operating range in applications. Further, the noise induced by the peripheral circuits can pass from one crossbar stage to another, causing errors in crossbar computations.

\subsection{Compensation techniques}

\paragraph{Devices} There are no universal techniques that compensate device-to-device variability. Each memristive device type is developed differently with various chemical and physical processes, which requires different approaches. For example, (1) in HfO$_x$ RRAM, the device variability was compensated by introducing an ultra-thin ALD-TiN buffer layer\cite{fang2018improvement}, (2) in Al$_2$O$_3$/TiO$_2$ (VMCO) RRAM, the use of non-filamentary RRAM reduces the impact\cite{govoreanu2013vacancy}, and (3) in SiO$_x$ RRAM, increasing the roughness of bottom electrodes reduces the device-to-device variability\cite{kenyon2019interplay}.

\paragraph{Peripheral circuits and programming} {The compensating schemes in literature are limited and largely an open problem. However, they are indirectly addressed via architectural modifications and system training, examples of which are listed below.}

\paragraph*{b.1) Architecture variants}
Modular crossbar array\cite{mikhailenko2018m} or tiled crossbar\cite{mountain2017memristor} is a method to split larger crossbars into smaller ones. Usually, large crossbars will have higher leakage and sneak path currents in their columns. The current errors can be reduced by performing MAC computations across multiple smaller crossbars. Simultaneously, it allows for a scalable configuration for large array processing applications.

The choice of the selector device\cite{zhou2014crossbar} is critical for the reliable operation of the crossbar. The common selectors used are transistors, diodes, nonlinear devices, and volatile switches. Both planar and vertical transistors could be used as a selector device depending on the 2D or 3D crossbar structure. Among diodes, Si p-n junctions, oxide/oxide heterojunction, and metal-oxide Schottky junctions are the most popular. The nonlinear devices include tunneling-based selections, complementary circuits, and mixed ionic electronic conduction, while volatile switches include phase transition, threshold, and short-retention switches.

Introducing redundancy in the crossbar node by parallelly combining memristors can help increase the stability and robustness of crossbar computation. Such nodes, known as superresolution nodes\cite{james2021analog}, can create a larger number of stable conductance levels per crossbar node for accurate analog computing.

\paragraph*{b.2) Training variants}

Training that involves updating the node conductance against a performance metric is useful in reducing errors resulting from crossbar variability. A variation-aware training often involves the addition of a ``penalty for variation" to conventional training for solving the conductance optimization process. This is practically useful for designing crossbar neural networks where the obtained weights deviate to new values due to crossbar variability and requires readjusting weights for optimal inference accuracy\cite{liu2015vortex}. 



Rather than an on-chip inference checking and training for the error, another approach is to perform offline training accounting for various crossbar variabilities \cite{krestinskaya2020automating}. This is useful for applications such as neural networks, where they need to be trained for variability and for reducing sensitivity to signal noise.

\section{Conclusion}

In conclusion, we note that the reliability of the memristor device and its ease of integration with matured technologies such as CMOS is crucial for realizing the majority of the practical applications with crossbars. Memristive devices need to have long retention and endurance, irrespective of the variability subjected in the crossbar for commercial use. Innovations in device material stack, crossbar architectures, and adaptive conductance adjustments through training effectively reduce the negative impacts of variability. In contrast, variability can also be useful for building systems requiring stochastic computing, including cryptography and stochastic neural networks.



%



\end{document}